**Spectroscopy of a Tunable Moiré System with a Correlated and Topological Flat Band**

Xiaomeng Liu[1*], Cheng-Li Chiu[1*], Jong Yeon Lee[2], Gelareh Farahi[1], Kenji Watanabe[3], Takashi Taniguchi[4], Ashvin Vishwanath[2], Ali Yazdani[1‡]

[1]*Joesph Henry Laboratories and Department of Physics, Princeton University, Princeton, NJ 08544, USA*
[2]*Department of Physics, Harvard University, Cambridge, Massachusetts 02138, USA*
[3]*Research Center for Functional Materials, National Institute for Materials Science, 1-1 Namiki, Tsukuba 305-0044, Japan*
[4]*International Center for Materials Nanoarchitectonics, National Institute for Materials Science, 1-1 Namiki, Tsukuba 305-0044, Japan*

\* These authors contributed equally to this work
‡ email: yazdani@princeton.edu

**Moiré superlattices created by the twisted stacking of two-dimensional crystalline monolayers can host electronic bands with flat energy dispersion in which interaction among electrons is strongly enhanced. These superlattices can also create non-trivial electronic band topologies making them a platform for study of many-body topological quantum states. Among the moiré systems realized to date, there are those predicted to have band structures and properties which can be controlled with a perpendicular electric field. The twisted double bilayer graphene (TDBG), where two Bernal bilayer graphene are stacked with a twist angle, is such a tunable moiré system, for which partial filling of its flat band, transport studies have found correlated insulating states[1–5]. Here we use gate-tuned scanning tunneling spectroscopy (GT-STS) to directly demonstrate the tunability of the band structure of TDBG with an electric field and to show spectroscopic signatures of both electronic correlations and topology for its flat band. Our spectroscopic experiments show excellent agreement with a continuum model of TDBG band structure and reveal signatures of a correlated insulator gap at partial filling of its isolated flat band. The topological properties of this flat band are probed with the application of a magnetic field, which leads to valley polarization and the splitting of Chern bands that respond strongly to the field with a large effective g-factor. Our experiments advance our understanding of the**

**properties of TDBG and set the stage for further investigations of correlation and topology in such tunable moiré systems.**

A moiré superlattice with flat electronic bands was first discovered by stacking two layers of graphene on top of each other at a precise angle[6–8]. The flat band in magic angle twisted bilayer graphene (MATBG) emerges from the interplay between interlayer hybridization and monolayer graphene' electronic structure and displays a plethora of correlated and topological phenomena. At partial band filling, MATBG shows a cascade of correlated insulating and superconducting phases, the underlying mechanisms of which are still being investigated[8–21]. A number of different Chern insulator phases, driven by either alignment of the MATBG with hexagonal BN or intrinsic strong correlations in such flat bands have also been found[22–26]. These discoveries motivate interest in other moiré systems and search for novel ways to create and control flat bands within them. A particularly promising moiré system is twisted double bilayer graphene (TDBG), in which the tunability of Bernal bilayer graphene band structure with an electric field can be exploited to create a tunable moiré system. Previous theoretical and experimental studies have demonstrated that application of an electric field can isolate a flat band with a bandwidth of 10~20meV[2,27–31] in TDBG. The electric field tunability makes it possible to realize this flat band in a wide range of twist angles (0.84°~1.53°) between the stacked bilayers, instead of at a specific angle as for the case of MATBG[2,3]. Corroborating this picture, previous transport experiments have shown that a robust correlated insulator appears in a confined electric field range at densities that correspond to half filling of a flat band in TDBG[1–5]. However, direct measurements of the band structure of this system and direct demonstration of its tunability, which underlies the interpretation of transport studies, has not been thus far performed.

We characterize the electronic structure of TDBG using scanning tunneling microscopy (STM) and spectroscopy as a function of carrier density and in the presence of a magnetic field. STM and gate-tuned scanning tunneling spectroscopy (GT-STS) have become a powerful tool to probe the properties of moiré systems, as its application to MATBG has provided critical information on the nature of flat bands in that system. STM experiments have not only confirmed the theoretical picture of the single particle band structure of MATBG but also have uncovered signatures of strong electronic correlation and topology at partial band filling of this system[16–19,24]. Applying these techniques to TDBG not only demonstrates its tunability but also provides spectroscopic signatures of correlation and topology of its flat bands that complements transport studies.

Figure 1a shows the schematic diagram of our device geometry which we use to probe the electronic properties of TDBG samples, in a home built ultra-high vacuum STM at a temperature T = 1.4K (see Methods for sample fabrication and STM measurement). Using STM imaging, we first obtain information on the structure and symmetry of the TDBG's moiré superlattice in our devices. Figure 1b shows an STM topography image, from which we can measure the periodicity and distortion of the moiré lattice to determine the twist angle of 1.48° and a very small hetero-strain of 0.1% for this TDBG sample. Higher resolution STM images of TDBG samples (Fig. 1b inset) show the topmost graphene layer's triangular atomic lattice, which signifies the sensitivity of our experiments to the electronic properties of this layer (out of four) in this multilayer system. STM images also reveal the breaking of $C_2$ symmetry—a salient feature of TDBG which is in contrast preserved in MATBG. This broken symmetry can be visualized by filtering the atomic lattice from STM images (Fig. 1b) and displaying the results in the moiré unit cell, as shown in Fig. 1c. From the STM topographic height in such images, we identify three different locations, ABBC, ABCA, and ABAB, within the unit cell that correspond to different stacking configurations of the atoms between the layers in our sample. While the locations corresponding to the AA stacking between the middle layers appear higher in STM images (Fig. 1b and 1c), there is also a measurable contrast in the height between AB or BA stacking locations as well. These locations are related by the $C_2$ symmetry of the moiré lattice, which is broken in TDBG. The broken $C_2$ symmetry is a distinguishing property of TDBG and is responsible for the gap between the conduction and the valence band at the Dirac points of its band structure as well as the unique band topology of TDBG.

Spectroscopy and visualization of the electronic states of TDBG over a wide range of energies reveal the basic band structure of this system and validates the continuum model that has been used to understand its properties. Figure 1e shows the STS measurement at the charge neutral point (CNP) that displays four distinct peaks corresponding to the van Hove singularities (vHs) of the four lowest energy bands of this system. The excellent agreement between these measurements and the density of states calculated by the continuum model (Fig. 1f, Supplementary Material) allows us to attribute the four identified bands to the first conduction and valence bands: C1, V1, and the second conduction and valence band: C2, V2 (Fig. 1d). Further corroboration for this assignment can be found by comparing the spatial dependence of the electronic states measured using STM conductance maps at energies corresponding to vHs of these bands and the maps of the electron distribution on the top graphene layer for the corresponding bands from the continuum model (Supplementary Material). Figure 2 shows an example of this comparison illustrating that the points of high density of states for V2, C1, and C2 bands are located on the ABBC moiré sites, while V1 displays a different pattern with its maximum density of states on ABCA sites. There is not only good agreement between the experimental results and the continuum model but also such maps clearly visualize the broken $C_2$ symmetry in TDBG by displaying the difference between the density of states at ABAB and ABCA sites.

The tunability of TDBG's band structure can be demonstrated by the measurements of its spectroscopic properties as we vary the gate voltage of the silicon gate $V_{SiG}$ in our devices. In similar previous experiments on gated devices with STM, such as those on MATBG (for example ref. [20]), varying $V_{SiG}$ allowed us to study the properties of such two-dimensional systems as a function of carrier concentration. However, for studies of TDBG, it is important to recognized $V_{SiG}$ tunes both the carrier concentration n and the electric field D simultaneously: $n = CV_{SiG}/e + n_0, D = CV_{SiG}/2 + D_0$, where C is the capacitance between TDBG and the silicon gate, $n_0$ and $D_0$ are the intrinsic doping and the built-in electric field. Measurements of the tunneling spectra display a rich array of features that shift systematically as a function of $V_{SiG}$, as shown in Figure 3a. First, we highlight that changes of doping in TDBG caused by the $V_{SiG}$ results in three clear jumps of the vHs marked by dashed white lines, indicating the Fermi energy is passing through a band gap. From these jumps, we identify the gate voltages corresponding to CNP ($V_{SiG}$ = -3.5V), full occupancy of the conduction band (72.5V), and full occupancy of the valence band (-79.5V). The appearance of CNP near zero gate voltage reveals that samples are not doped by impurities or by a significant work function difference between the sample and the STM tip. Furthermore, we find the gate voltage ranges required to occupy the conduction band and the valance band are identical: $\Delta V_{SiG}$ = 76V, which is also consistent with the carrier density needed to fill a moiré band of the 1.48° TDBG: $n_S$ = 5.08*$10^{12}$cm$^{-2}$ based on a parallel plate capacitance model (see Methods). These observations indicate that our sample is pristine, and our measurements are free from artifacts of tip-induced band bending, which was a concern in early STM studies of MATBG[16–19].

The structure of TDBG bands, especially those of the C1 and the V1 bands, undergoes drastic changes in spectroscopic measurements as we tune the electric field D in our device by changing $V_{SiG}$. Before increasing D, at the CNP, there are large band gaps between V2 and V1 ($\Delta_{-n_s}$), C1 and C2 ($\Delta_{n_s}$), with $\Delta_{-n_s} < \Delta_{n_s}$ (Fig. 1e). Figures 3 (a, c, and d) show that increasing the magnitude of the electric field $|D|$ with increasing $|V_{SiG}|$ results in reduction of the magnitudes of these energy gap, with $\Delta_{-n_s}$ being almost eliminated by large electric fields, as displayed at the top and bottom portions of Fig. 3a. These gaps reduce in a symmetric manner toward the positive and negative gate voltages, signifying that the built-in electric $D_0$ is close to zero. These findings are consistent with previous transport studies of TDBG and show that the band structure of TDBG can be tuned with an electric field. In addition to these gaps, we identify a smaller gap between C1 and V1 bands at charge neutrality $\Delta_{CNP}$, which behaves in an asymmetric fashion with respect to the sign of the $V_{SiG}$. At positive values of $V_{SiG}$, $\Delta_{CNP}$ is first reduced in magnitude to zero but it reopens again with increasing $V_{SiG}$ (Fig. 3a, d), in agreement with transport experiments at this twist angle[2]. In contrast, $\Delta_{CNP}$ monotonously increases in magnitude as $V_{SiG}$ is made more negative (Fig. 3a, c). This asymmetry with respect to $V_{SiG}$ is due to the polarizing effect of the electric field, which induces changes in the spatial distribution of electronic states across the different atomic layers of TDBG. Our experiments are sensitive to this asymmetry because they are sensitive to the electronic properties of the topmost graphene layer in TDBG. In fact, comparing our experimental results of the gate-tuned measurements of tunneling spectra to the continuum model (Supplementary Material), we find the calculated density of states for the top graphene layer (Fig. 3b) from this model shows a remarkable resemblance to our data (Fig. 3a), including the observed asymmetric behavior with respect to $V_{SiG}$. In these calculations, the tuning of electric field D is modeled by the adjustable parameter U, which is the onsite potential difference between the top and bottom graphene layer. Furthermore, the continuum model allows us to understand why we do not observe the closing of $\Delta_{CNP}$ at negative $V_{SiG}$ – it actually closes, but the signature appears only in the density of states of the bottom layer, which is not being probed by our experiment. One might notice a slight discrepancy between the model and the experiment, that there is a small overlap between V1 and C1 in the calculated band structure (Fig. 1d and Fig. 3b), in contrast to the clear CNP gap seen in our experiment and transport experiment[2]. In fact, with our choice of model parameters, a CNP gap at D=0 and its closing and opening with the application of D is predicted for a twist angle larger than 1.52°, and it is likely that the small overlap between C1 and V1 bands at the angle of 1.48° will be gapped by electron-electron interactions or coupling with the substrate. Lastly, we find that the bandwidth of C1 in both experiment and model calculation is smaller than that of V1, which explains why correlated insulators have only been discovered in the conduction band and establishes the setting for the correlation signatures to emerge in our experiments.

The influence of strong electronic correlation in the TDBG system can be examined in higher resolution gate-tuned spectroscopy measurements of the C1 and V1 bands. Figure 4a shows such measurements in which we find a number of different features as the occupation of the C1 band is tuned. First, we note the abrupt jump in vHs of C1 at CNP due to the CNP gap, the closing of which can be identified near $V_{SiG}$~ 20V in this higher resolution measurement. Second, as the C1 is being occupied, we also observe a sudden increase in the broadening of the vHs of the fully occupied V1 band, reminiscent of similar features seen in such measurements of MATBG and understood as a results of strong correlation in that system[17]. Third and most intriguingly, increasing the occupation of C1 to half filling (at $V_{SiG}$=34.5V), where transport studies have uncovered a correlated insulator, we observe a sudden jump of the vHs of this band (Fig. 4a-c, Figs. S1-S2). This jump sometimes appears together with a suppression of the density of states at the Fermi level, the measurements of which we find to be sensitive to the location of the tip and also likely to its work function (inset Fig. 4a and discussion on Figs. S1-S2 in Methods). Examining the behavior of vHs more closely by tracking its location in energy at different $V_{SiG}$, we extract the magnitude of its jumps Δ at different twist angle areas. Consistent with correlation gaps extracted from transport studies at half filling of the C1 band, we find the magnitude of Δ increases from 0.38meV to 1.3meV as the twist angle decreases from 1.48° to 1.43°. The observation of jump-like features in our data is not expected from a simple insulating gap, such as observed single particle gaps observed at CNP and full occupancy of C1 and V1 bands. Its shape is however reminiscent of similar measurements on MATBG near every quarter fillings of that system[20,21] and requires further measurements and theoretical analysis.

Lastly, we show that topology of the band structure of the TDBG band can be revealed by examining its spectroscopic properties in a magnetic field. In the presence of an electric field *D*, the C1 band of TDBG in our theoretical model carries a topological valley Chern number $C_V$ = 2, meaning Chern number C=2 in one valley and C=-2 in the opposite valley. Since a Chern band generates orbital magnetic moments, it would couple strongly to a perpendicular magnetic field. In Fig. 5a, we show a gate-dependent STS measurement under a perpendicular magnetic field of 4T. Compared with zero field data (Fig. 4a), the most salient feature is the splitting of the vHs of the C1 band. This behavior broadly agrees with the expectation of a topological valley Chern band. Since the two valleys of the C1 band carry opposite Chern numbers and orbital g-factors, they shift in energy under a magnetic field in the opposite directions by $\pm g\mu_B B$, where $\mu_B$ is the Bohr magneton and g is the magnitude of the orbital g-factor (Fig. 5d, e). While in general g-factors depend on crystal momentum, here for simplicity we discuss the average g-factor[28]. Thus, the magnetic field split the valley-degenerate vHs of the C1 band under zero magnetic field into two vHs of different valleys separated by $2g\mu_B B$. Extracting the separation between peaks at different magnetic fields (Fig. 5b and Fig. S3), we find they collapse onto lines with very large g-factors of g=13 and 21 for half-filling and three-quarter filling (Fig. 5c). From the continuum model calculation, g-factor averaged in the mini-Brillion zone is between 6 and 7 for most U (Supplementary Material), which is a factor of two times smaller than the experimental value. However, the g-factor calculation sensitively depends on the energy landscape within the narrow C1 band, which may be altered by interactions that are not accounted for in the single particle picture. Without knowing the exact shape of the band, a simple picture below may provide a rough estimate of the g-factor amplitude for a general $C_v$=2 band. Since the CNP gap can be obtained by emptying of the C1 band from the trivial $n_S$ gap, the CNP gap possesses a valley Chern number that is opposite of the C1 band. At the $K_1$ ($K_2$) valley, the CNP gap would shift in energy from the bottom (top) edge of C1 (V1) band at zero magnetic field to the top (bottom) of C1 (V1) band at half flux quantum per moire unit cell[32]. If the vHs of C1 roughly shift along with the CNP gap, and if we assume the CNP gap is much smaller than the bandwidth of C1 and V1, the splitting between vHs of the C1 band in two valleys produced by the field would be on the order of $B/B_{1/2\Phi} * (w_c + w_v)$, where $B_{1/2\Phi} = 26.3\ T$ is the magnetic field of half flux quantum per moire unit cell and $w_c$ ($w_v$) is the bandwidth of the C1 (V1) band. Assuming the bandwidth of C1 and V1 are on the order of 20meV, this formula produces a g-factor g=13, roughly agreeing with our observation. Beside the splitting, we also resolve a discontinuity of the vHs just below $V_{SiG} = 20V$, where the CNP gap collapses in Fig. 4a, and a jump of vHs near three-quarter filling of the C1 band. These features and other topological signatures of this system that are revealed in the present of a magnetic field[32], can be explored in future studies of this system.

## Methods

**Sample preparation.** TDBG samples are stacked on a hexagonal Boron Nitride (hBN) layer on SiO$_2$/silicon substrate and electrically contacted with gold electrodes. The samples are fabricated by tearing and stacking an exfoliated Bernal stacked bilayer graphene. We fabricated our samples in similar ways to published STM results on MATBG[17,20]. Our pickup stamp is made from a polyvinyl alcohol (PVA) coated transparent tape, which cover a polydimethylsiloxane (PDMS) block on a glass slide handle. Using the PVA stamp, we first pick up an exfoliated hBN flake. We then use the hBN to tear and pickup half of an exfoliated Bernal bilayer graphene. After rotating the stage about 1.3°, we pick up the remaining bilayer graphene flake. To flip the stack upside down, we transfer the stack onto a new PDMS block and dissolve PVA with water to detach the original pickup stamp. After cleaning in water, we drop the flipped stack from the PDMS block to a SiO$_2$/Si substrate with prepatterned gold contact. Once the device is made, we briefly dip it into water to clean PVA residues and then annealed it in UHV at 180°C overnight.

**STM measurements.** The experiment is done in a home built UHV STM with (6,1,1) vector magnet operating at T=1.4K. The measurements are performed with a tungsten tip prepared and characterized on a Cu(111) single crystal. Through controlled indentation, we shape the tip until its poke mark is confined and its spectrum featuring the Cu(111) surface state at the right energy. We then locate the TDBG sample with the capacitance guiding technique[33]. The topography is taken under constant current mode, and the differential conductance is acquired by standard lock-in method with AC modulation at 4kHz.

**Estimated silicon gate voltage required to fill a moiré band.** The moiré unit cell area for our 1.48° TDBG is A = 78.7nm$^2$. Thus, the electron density corresponds to full filling is n$_S$ = 4/A = 5.08*10$^{12}$cm$^{-2}$. In our experiment, the SiO$_2$ sickness is ~285nm and the hBN thickness is estimated to be ~30nm. Assuming the dielectric constants of SiO$_2$ and hBN are both $\epsilon_r = 4$, the gate voltage need to dope the TDBG from CNP to n$_S$ is then $\Delta V_{SiG} = 72.4V$, similar to the experimental value of 76V.

**Discussion on Fig. S1-S2.**

In Fig. 4a-c and Fig. S1-S2 we show several GT-STS measurements that provide information on the variation of such measurements at different areas of the same device as well as different locations within the moiré unit cell. Figure S1 highlights that the signatures of correlations in the broadening and the visibility of the gap at the Fermi level in C1 can be different when measured on different locations and with likely different tip conditions. In Fig. S2, we include more GT-STS measurements presented together with STM topographs that show areas 1-4. The location of the corresponding measurements within the unit cell are marked on these topographs (Fig. S2a-d). These measurements (Fig. S1 and same data reproduced in Fig. S2h) suggest that the correlation gap at the Fermi level and broadening of C1 are enhanced on the ABBC site (Area 4, Fig. S2d). This observation is consistent with enhancement of the density of state of the C1 band at these location as visualized in Fig. 2 and discussed in the main text. Fig. 2S also shows areas 1-3 and corresponding DT-STS used in Fig. 4a-c, which produce the analysis in Fig. 4d-f.

## Acknowledgements

We thank Andrei Bernevig for helpful discussions. A.V., J.Y.L. and X.L. would like to thank Zeyu Hao, Eslam Khalaf, Shang Liu and Philip Kim for an earlier collaboration on TDBG. This work was primarily supported by the Gordon and Betty Moore Foundation's EPiQS initiative grants GBMF4530, GBMF9469, and DOE-BES grant DE-FG02-07ER46419 to A.Y. Other support for the experimental work was provided by NSF-MRSEC through the Princeton Center for Complex Materials NSF-DMR-1420541, NSF-DMR-1904442, ExxonMobil through the Andlinger Center for Energy and the Environment at Princeton, and the Princeton Catalysis Initiative. A.V. and J.Y.L were supported by a Simons Investigator fellowship. K.W. and T.T. acknowledge support from the Elemental Strategy Initiative conducted by the MEXT, Japan, grant JPMXP0112101001, JSPS KAKENHI grant JP20H00354, and the CREST (JPMJCR15F3), JST.

## Author Contributions

X.L., C.C. and A.Y. designed the experiment. X.L. and C.C. fabricated the samples. X.L., C.C. and G.F. performed the STM measurements and analyzed the data. J.Y.L., A.V. and X.L. conducted the theoretical calculations. K.W. and T.T. provided hBN crystals. X.L., C.C., J.Y.L and A.Y. wrote the paper with input from all authors.

**Competing Financial Interests**

The authors declare no competing financial interests.

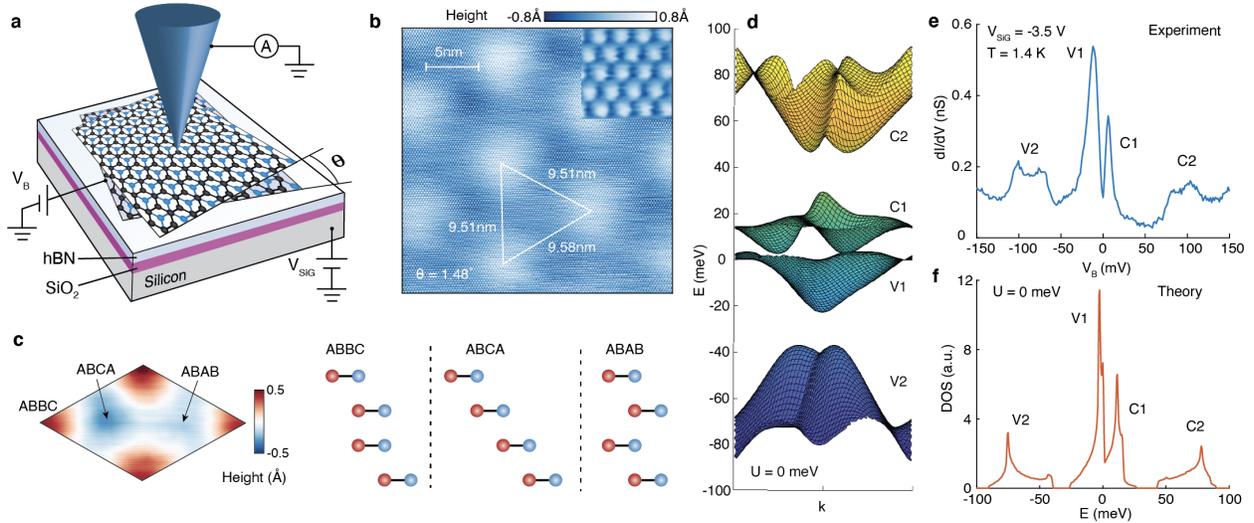

**Figure 1 | STM on TDBG. a,** Schematic of the STM measurement setup on TDBG devices. **b,** STM topograph of TDBG, obtained with bias setpoint $V_{set}$ = -400mV, current setpoint $I_{set}$ = 20pA and gate voltage $V_{SiG}$ = -88V. Top right inset shows magnified image (1nm size) with top graphene layer lattice in a Bernal bilayer. **c,** Filtered topograph removing atomic lattices to emphasize different stacking orders of TDBG plotted in the moiré unit cell. The colored circles on the right represent carbon atoms on A and B sublattices of each layer. **d,** Calculated band structure of TDBG with a twist angle of $\theta$ = 1.48° under zero potential different between the top and the bottom graphene layers (U = 0). **e,** Tunneling spectrum of 1.48° TDBG device at charge neutrality, measured with $V_{set}$ = -400mV, $I_{set}$ = 400pA and AC modulation of $V_{mod}$ = 3mV. **f,** Density of state on the top graphene layer calculated by a continuum model under zero electric field.

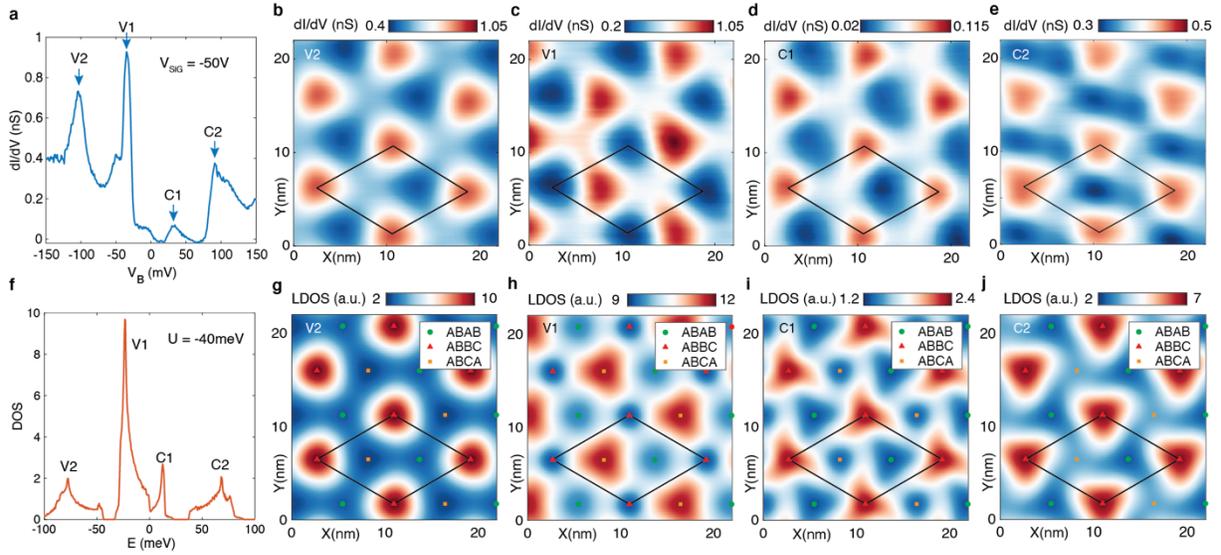

**Figure 2 | Visualizing Electronic States of TDBG. a,** Tunneling spectrum at $V_{SiG}$ = -50V and T = 1.4K, measured with $V_{mod}$ = 5mV, $V_{set}$ = -200mV, $I_{set}$ = 400pA. **f,** Calculated DOS with parameter U = -40meV, which is the onsite potential difference between the top and bottom graphene layer. **b-e,** Conductance maps for V2, V1, C1 and C2 bands, taken at T = 1.4K, $V_{SiG}$ = -50V, $V_{mod}$ = 5mV. The voltage setpoint of each map is set to the spectrum peaks in **a** at $V_{set}$ = -103mV, -35mV, 32mV and 91mV, while the current setpoint is chosen to keep the tip at roughly the same height as **a**. The plotted data was smoothed with a 0.8nm radius filter. The diamond shapes represent a moiré unit cell. **g-j,** Calculated electron density distribution on the top graphene layer for the four bands. U = -40meV is used for this calculation. Shown electron densities are summed over each band (Supplementary Material), which appears to capture the data under these STM setup conditions. The diamonds match with the ones in **b-e**. Different stacking orders are shown by different colored symbols.

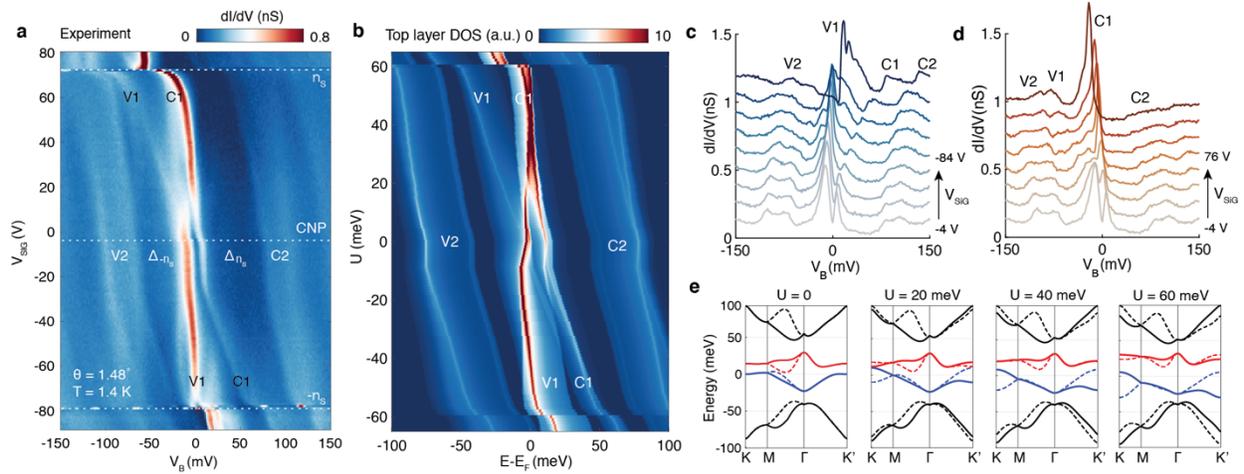

**Figure 3 | Tunable band structure of TDBG. a,** Differential tunneling conductance as a function of bias voltage $V_B$ and silicon gate $V_{SiG}$ measured near the ABAB stacking configuration, with $V_{set}$ = -400mV, $I_{set}$ = 400pA and AC modulation of $V_{mod}$ = 3mV. The white dashed lines mark the charge neutral point (CNP) and full filling of C1 ($n_S$) and V1 (-$n_S$) bands. **b,** Calculated top graphene layer density of state as a function of energy respect to Fermi energy $E-E_F$ and onsite potential difference between the top and bottom graphene layer U. **c, d,** line spectra from CNP to full hole (**c**) and electron (**d**) filling. **e,** Band structure at different U in the mini-Brillouin zone, red (blue) lines represent the first conduction (valence band). The solid and dashed lines correspond to two different valleys $K_1$ and $K_2$ of the Brillouin zone of the graphene lattice.

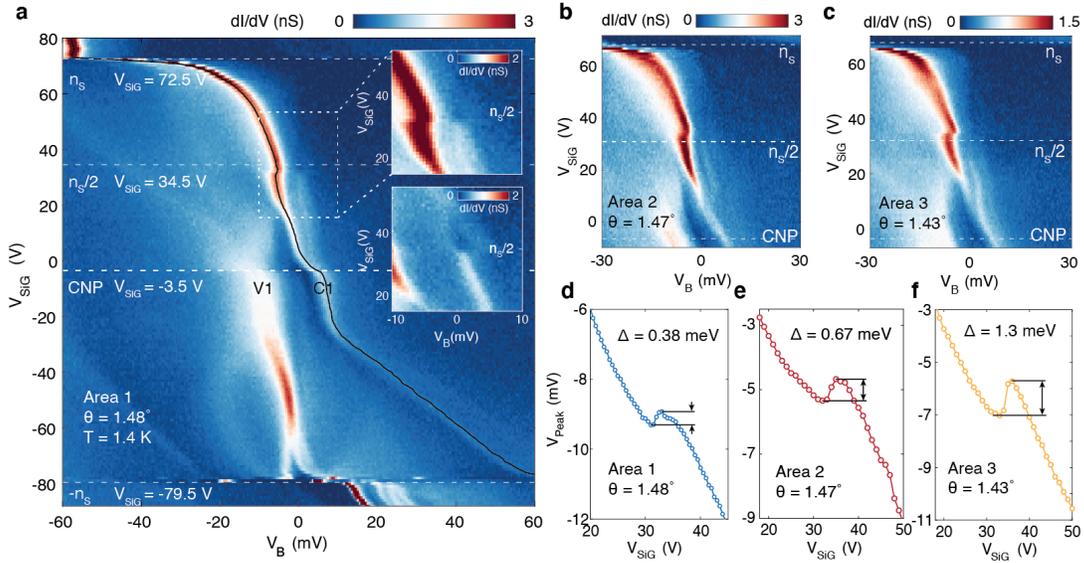

**Figure 4 | Evidence for correlated insulating states at half filling of the conduction band. a,** High resolution gate-dependent spectra in the first area with $\theta = 1.48°$, taken near the ABAB stacking configuration with $V_{set}$ = -400mV, $I_{set}$ = 800pA and $V_{mod}$ = 0.5mV. There is a clear jump of vHs near the half filling of the conduction band. Top inset, zoom-in around the half filling and zero bias, revealing a small dip in tunneling conductance at zero bias near half filling. Bottom inset, a zoom-in of Fig. S1, showing a more pronounced gap near half filling. **b, c,** High resolution maps showing similar jumps of vHs in different areas of the sample with twist angles 1.47° and 1.43°, also taken with $V_{mod}$ = 0.5mV. **d-e,** Traces of vHs peaks over silicon gate voltages in three different areas. The magnitude of the jump Δ in each area is extracted.

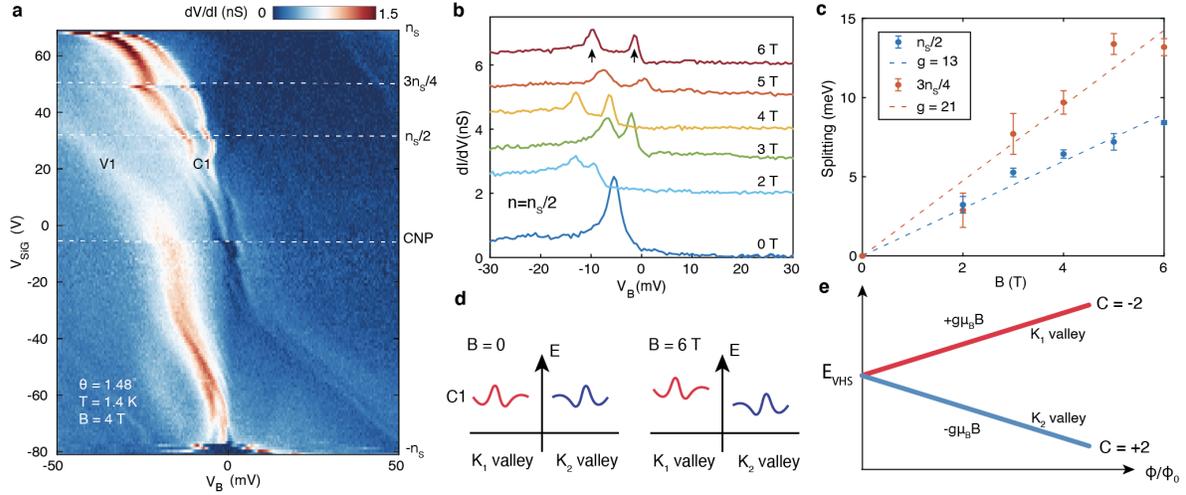

**Figure 5 | Splitting of vHs under perpendicular magnetic fields. a,** STS measurements between -$n_S$ and $n_S$ under a perpendicular magnetic field of 4T, taken with $V_{set}$ = -400mV, $I_{set}$ = 800pA and $V_{mod}$ = 0.5mV. **b,** Tunneling spectra at half filling of the C1 band under various perpendicular magnetic fields. Traces are offset for clarity. **c,** Splitting of the C1 vHs as a function of perpendicular magnetic field B at half filling (blue symbols) and three-quarter filling (red symbols), obtained from Gaussian fitting of the split peaks. The dashed lines are the linear fit of the splitting, corresponding to $\Delta E = 2g\mu_B B$ with g-factor noted in the legend. **d, e,** schematic of C1 band's response to magnetic fields. The two valleys (red and blue) shift in opposite energy directions due to their opposite valley Chern numbers.

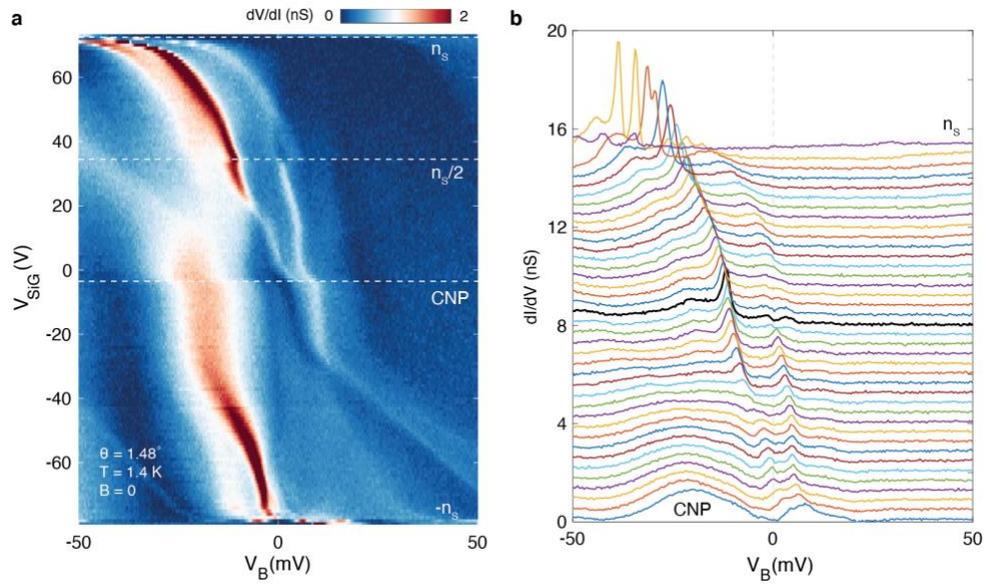

**Figure S1 | Gate-dependent spectra showing gaps at the Fermi level near half-filling of C1.**
**a,** STS measurements between -n$_s$ and n$_s$, taken at ABBC stacking area (See Fig. S2, area 4), with V$_{set}$ = -350mV, I$_{set}$ = 800pA and V$_{mod}$ = 0.5mV. **b,** Line traces of the data presented in a. The curves are shifted vertically for clarity. The bottom curve is at charge neural point and the top curve is at n$_s$. The trace at half filling is highlighted by the thicker black curve. We note a single peak split into two peaks at half filling.

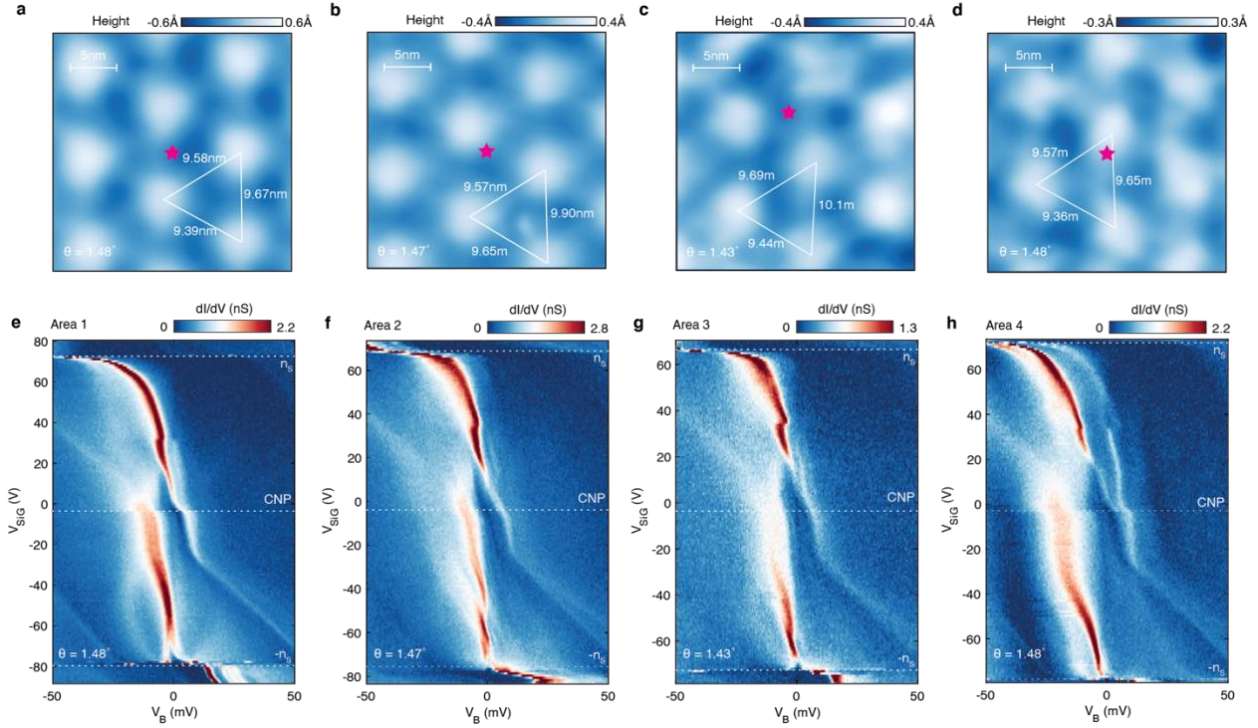

**Figure S2 | Topographs and gate-dependent STS measurements at various locations. a-d,** STM topographs of TDBG taken at area 1, 2, 3 and 4 respectively. The bias set point $V_{set}$ and current set point $I_{set}$ are **a,** $V_{set}$ = -150mV, $I_{set}$ = 200pA. **b,** $V_{set}$ = -400mV, $I_{set}$ = 40pA. **c,** $V_{set}$ = -400mV, $I_{set}$ = 50pA. **d,** $V_{set}$ = -400mV, $I_{set}$ = 70pA. **e-h,** gate-dependent STS measurements taken at area 1, 2, 3 and 4 respectively. The red stars mark the location where the gate-dependent STS is taken. The white dashed lines mark the charge neutral point (CNP) and full filling of C1 ($n_S$) and V1 (-$n_S$) bands. All the data are taken with AC modulation $V_{mod}$ = 0.5mV. The bias set point $V_{set}$ and current set point $I_{set}$ are **e,** $V_{set}$ = -400mV, $I_{set}$ = 800pA. **f,** $V_{set}$ = -300mV, $I_{set}$ = 800pA. **g,** $V_{set}$ = -400mV, $I_{set}$ = 800pA. **h,** $V_{set}$ = -400mV, $I_{set}$ = 800pA.

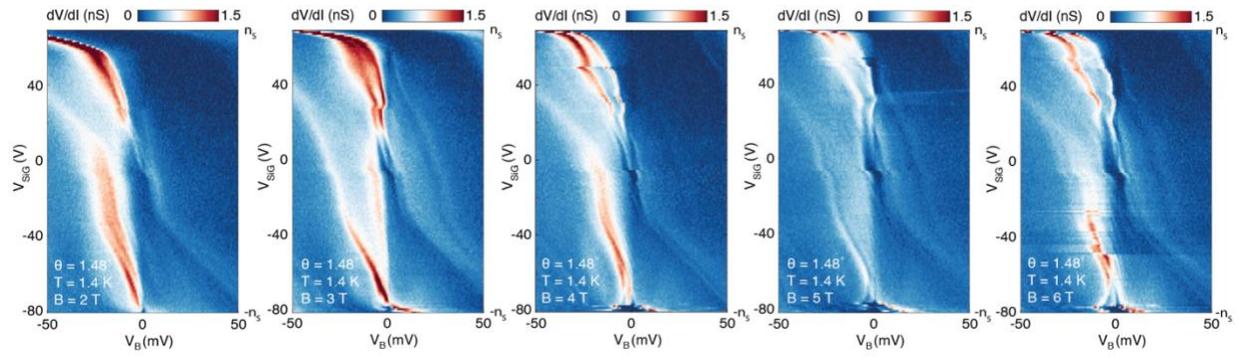

**Figure S3 | Gate-dependent STS measurements under various magnetic fields.** STS measurements between $-n_s$ and $n_s$ under five different magnetic fields. All data shown here are taken with voltage set point $V_{set}$ = -400mV, current set point $I_{set}$ = 800pA and AC modulation $V_{mod}$ = 0.5mV. The line-cuts and splitting shown in Fig. 5 are extracted from these data.

# Supplementary Material
(Dated: July 17, 2020)

## I. CONTINUUM MODEL CALCULATION

In this section, we elaborate how the density of states plot corresponding to the STM experiment can be calculated. We used the continuum model for TDBG as in the published work ([1]). The parameters used in our calculations are the following:

$$(\gamma_0, \gamma_1, \gamma_3, \gamma_4, \Delta, w_0, w_1) = (2700, 361, 283, 138, 15, 88, 110) \text{ meV}. \quad (1)$$

For given model parameters, we calculated band structures for different set of twist angles $\theta$ and potential difference between layers $U$. The density of state on the top graphene layer is extracted by considering the wavefunction distribution of the top layer. Each eigenstate $\Psi_{n,\tau,\sigma}(\bm{k})$ is labelled by the band index $n$, valley index $\tau$, spin index $\sigma$, and momenta $\bm{k}$, consisting of eight components – four layers, and two sublattice sites $A$ and $B$ for each layer. To obtain the density of states in the top-most layer (for each valley and spin), we evaluated the following:

$$\rho_{\tau,\sigma}^{\text{top}}(E)\Delta E = \sum_{E_{\bm{k}} \in (E, E+\Delta E)} \left( \left|\Psi_{n,\tau,\sigma}^{A1}(\bm{k})\right|^2 + \left|\Psi_{n,\tau,\sigma}^{B1}(\bm{k})\right|^2 \right) \quad (2)$$

where $\Delta E$ is the energy resolution. To compare with the STM experiment in Fig. 3, we must shift the calculated density of state in energy so that it is plotted as a function of the bias voltage $V_B$, which is equivalent to the energy relative to the Fermi energy $E - E_F$. Note that the Fermi energy is determined by the electron density $n_e$ and the band structure. Therefore, we have to know the electron density $n_e$ at each value of $U$.

In the experiment, the silicon gate voltage tunes both the carrier density and electric field simultaneously. It is fair to assume that both quantities change linearly in the gate voltage. Based on the experimental observation of the gap closings between the first and second valence bands, we estimated the correspondence between the continuum model parameter $U$ and experimental gate voltage $V_{\text{SiG}}$, which in turn is proportional to the electron density $n_e$. As a result, in the calculations, we use the following quantitative relation between the electron density $n$ and $U$:

$$n_e = n_s \cdot \frac{U}{60 \text{ meV}}, \quad (3)$$

where $n_s$ corresponds to the full-filling electron density, which is four electrons per Moire unit cell.

With a given electron density $n_e$ and band structure at certain $U$, it is straightforward to evaluate the Fermi energy. Here we used the mesh size of $80 \times 80$ in the momentum space.

## II. REAL SPACE DENSITY OF STATES

In the tight-binding model, there is no real space distribution of real space density for a particular state. However, if each unit cell contains multiple sites, there can be a non-uniform distribution of electrons within the unit cell. Similar thing happens for the continuum model where each wavefunction is the linear combination of Fourier components which differ by Moire reciprocal lattice vectors. In numerics, each eigenvector corresponds to the following vector

$$\Psi(\bm{k}) = (u_{\bm{k}}(\bm{G}_1), u_{\bm{k}}(\bm{G}_2), u_{\bm{k}}(\bm{G}_3), ..., u_{\bm{k}}(\bm{G}_i), ...), \quad (4)$$

where $G_i$ is the set of Moire reciprocal lattice vectors. In practice, we truncate at certain point and keep only a finite number of Fourier components. Its spatial distribution within the Moire unit cell is simply calculated as the following:

$$\rho_{\bm{k}}(\bm{r}) = |\Psi_{\bm{k}}(\bm{r})|^2 = \left| \sum_{\bm{G}} e^{i(\bm{k}+\bm{G})\cdot \bm{r}} u_{\bm{k}}(\bm{G}) \right|^2. \quad (5)$$

Finally, the real space density plot can be summed over the energy window we are interested in.

## III. CHERN NUMBER AND $g$-FACTOR

In the main text, we remarked that the observed large $g$-factor is related to the band carrying non-zero Chern number. This can be understood by examining how they are defined. A Chern number is defined as the integration of the Berry curvature, which is defined as the following:

$$C_n = \frac{1}{2\pi} \int d^2k \, \Omega_n(\boldsymbol{k}), \qquad \Omega_n(\boldsymbol{k}) = i \sum_{m \neq n} \frac{\langle n| \partial_{k_x} H_{\boldsymbol{k}} |m\rangle - \text{h.c.}}{(E_n - E_m)^2}, \tag{6}$$

where $\Omega_n(\boldsymbol{k})$ is the berry curvature of $n$-th band at momentum $\boldsymbol{k}$. This definition is consistent with the definition where the electron's Landau level carries $C = +1$. Note that $\sigma_{yx} \propto C$ in this case. The formula for berry curvature is similar to the orbital $g$-factor defined as the following [1]:

$$g_n(\boldsymbol{k}) = \frac{2m_e}{\hbar} i \sum_{m \neq n} \frac{\langle n| \partial_{k_x} H_{\boldsymbol{k}} |m\rangle - \text{h.c.}}{E_n - E_m}, \tag{7}$$

where $m_e$ is the free electron mass. The main difference between $\Omega_n(\boldsymbol{k})$ and $g_n(\boldsymbol{k})$ is the denominator.

For simplicity, consider a two-band model with $E_1(\boldsymbol{k}) > E_2(\boldsymbol{k})$. There, it is easy to notice that $\Omega_1(\boldsymbol{k}) = -\Omega_2(\boldsymbol{k})$. Furthermore, we can notice that

$$g_1(\boldsymbol{k}) = g_2(\boldsymbol{k}) = \frac{2m_e}{\hbar}(E_1 - E_2) \cdot \Omega_1(\boldsymbol{k}). \tag{8}$$

Therefore, we can deduce the relation between averaged $g$-factor and Chern number as the following

$$g_1^{\text{ave}} = g_2^{\text{ave}} \sim \frac{2m_e}{\hbar}(E_1 - E_2) \cdot \frac{2\pi C_1}{A_{\text{BZ}}} \tag{9}$$

where $A_{\text{BZ}}$ is the area of the Brillouin zone. Of course, this would not hold if each band is so dispersive that the gap $E_1(\boldsymbol{k}) - E_2(\boldsymbol{k})$ fluctuates significantly, which gives a correction proportional to the manitude of the fluctuation. However, assuming two bands are well separated and relatively flat, one can associate Chern number and $g$-factor averaged over the Brillouin zone.

In twisted materials, know that $g$-factors satisfy the following relationship between opposite valleys:

$$g_{n,+}(\boldsymbol{k}) = -g_{n,-}(-\boldsymbol{k}) \quad \Rightarrow \quad g_{n,+}^{\text{ave}} = -g_{n,-}^{\text{ave}}. \tag{10}$$

This implies that under the strong perpendicular magnetic field, we would observe the splitting of van-Hove singularities originated from different valleys. In our TDBG system, we notice that the first conduction and valence bands are well isolated from the other bands. Over the range of $U$ values we calculated, the direct gap between first and second conduction (valence) bands is factor of 2 to 4 times larger than the direct gap between first conduction and valence bands. Therefore, we can associate the observation of large $g$-factor and the band carryring non-zero Chern number.

---